\documentclass[letterpaper]{article}
\usepackage{nips,times}
\usepackage[dcucite]{harvard}
\usepackage{graphicx}
\usepackage{subfigure}
\usepackage{url}
\usepackage{common}

\begin{document}

\title{TextLuas: Tracking and Visualizing Document and Term Clusters in Dynamic Text Data}


\author{
Derek Greene \\
Insight Centre for Data Analytics\\
University College Dublin, Ireland
\And
Daniel Archambault \\
Department of Computer Science\\
Swansea University, UK
\And
V\'{a}clav Bel\'{a}k \\
Insight Centre for Data Analytics\\
NUI Galway, Ireland
\And
P\'{a}draig Cunningham \\
Insight Centre for Data Analytics\\
University College Dublin, Ireland
}
 
\maketitle
\begin{abstract}
For large volumes of text data collected over time, a key knowledge discovery task is identifying and tracking clusters. These clusters may correspond to emerging themes, popular topics, or breaking news stories in a corpus. 
Therefore, recently there has been increased interest in the problem of clustering dynamic data. However, there exists little support for the interactive exploration of the output of these analysis techniques, particularly in cases where researchers wish to simultaneously explore both the change in cluster structure over time and the change in the textual content associated with clusters.
In this paper, we propose a model for tracking dynamic clusters characterized by the evolutionary events of each cluster.  Motivated by this model, the TextLuas system provides an implementation for tracking these dynamic clusters and visualizing their evolution using a metro map metaphor.  To provide overviews of cluster content, we adapt the tag cloud representation to the dynamic clustering scenario. We demonstrate the TextLuas system on two different text corpora, where they are shown to elucidate the evolution of key themes. We also describe how TextLuas was applied to a problem in bibliographic network research.

\end{abstract}


\section{Introduction}
\label{sec:intro}

In many domains, where the data has a temporal aspect, it will be of interest to analyze the formation and evolution of patterns in the data over time. For instance, researchers may be interested in tracking evolving communities of social network users, such as clusters of members with shared interests on social media sites. 
In the case of online news sources, each producing a substantial volume of articles on a daily basis, it will often be useful to chart the progress of individual news stories as they develop over time.

A number of authors have considered the problem of tracking clusters over time \cite{palla07quant,greene10tracking}. Often these techniques involve dividing continuous data into a series of successive ``time steps'', which can be analyzed in turn. However, these techniques have largely focused on the problem of finding dynamic communities in social networks. In addition, relatively little focus has been given to the problem of interactively visualizing cluster evolution -- instead researchers have mostly relied on non-interactive diagrams of cluster evolution generated either manually or semi-automatically on small data sets \cite{palla07quant,tanti07framework}.   In initial work on increasing the scalability of such methods, \citeasnoun{rosvall10change} have emphasized the evolution of clusters over time.  However, such techniques do not consider a key aspect of many dynamic data sets -- the simultaneous change in group memberships \textit{and} in the textual content associated with clusters (\eg frequent terms appearing in user-generated posts, popular tags appearing in social bookmarking sites).

In this paper, we seek to address the absence of a solution in this space, through the development of a methodology for simultaneously exploring both cluster evolution and content shifts in dynamic text data sets, where clusters correspond to themes or topics that develop over time. We propose that the dynamic data analysis task can be framed as the problem of tracking clusters of nodes in multiple related bipartite graphs. We present a model for tracking clusters over multiple time steps in a dynamic network, where the ``life-cycle'' of each cluster is characterized by a \emph{timeline} indicating significant events. Using this model, we have developed a visualization system for exploring both evolving cluster timelines and the changing nature of their content.  This system illustrates the relationship between dynamic clusters using a metro map-style layout, and the content contained in these clusters is presented in the form of aggregated tag clouds that are indicative of cluster content over a specific time period.  The system, TextLuas, derives its name from the \emph{Luas} tram service in Dublin, Ireland.  

The remainder of the paper is structured as follows. In the next section, we provide a summary of relevant existing work pertaining to cluster analysis and visualizing dynamically evolving text and graphs. In \refsec{sec:methods}, we outline the proposed tracking model and provide a detailed description of the term cluster tracking method. While the proposed system can potentially be used independently of any specific clustering algorithm, in \refsec{sec:cocluster} we discuss a suitable choice of algorithm which can be applied to identify co-clusterings in individual time steps, by taking into account both current and historic data. In \refsec{sec:viz}, we describe our visualization techniques.  
Results of our proposed visualization system on two real-world text collections, from economic news sources and a Web 2.0 social bookmarking portal, are discussed in \refsec{sec:eval}. Subsequently in \refsec{sec:UserFeedback} we provide user feedback relating to the application of TextLuas by researchers interested in exploring bibliographic networks. The paper concludes with suggestions for plans for future work in the area of dynamic data analysis.

\section{Related Work}
\label{sec:related}

\subsection{Cluster Analysis}
\label{sec:reldyn}

\subsubsection{Document Clustering}
\label{sec:reldoc}
A wide range of algorithms have been proposed for the unsupervised exploration of text corpora \cite{ghosh03book}. A common approach has been to apply partitional algorithms, such as $k$-means or one of its many variants, to produce a disjoint partition of a document collection \cite{dhillon01sphere}. 
An alternative approach to the document clustering problem is to ``co-cluster'' documents and terms simultaneously. \citeasnoun{dhillon01cocluster} proposed representing a document collection as a weighted bipartite graph -- the two node types in the graph correspond to terms and documents respectively, while edges between two nodes indicate how frequently a term occurs in a given document. The document clustering problem then involves partitioning the bipartite graph, which can be achieved efficiently by via a spectral approximation to the optimal normalized cut of the graph. Specifically, \citeasnoun{dhillon01cocluster} proposed computing a truncated singular value decomposition (SVD) of a suitably normalized term-document matrix, constructing an embedding of both terms and documents, and applying $k$-means to this embedding to produce a simultaneous $k$-way partitioning of both documents and terms. 

\subsubsection{Evolutionary Clustering}
\label{sec:reldyn1}
The general problem of identifying clusters in dynamic data has been studied by a number of authors. Early work on the unsupervised analysis of temporal data focused on the problems of topic tracking and event detection in document collections \cite{yang98study}. More recently, \citeasnoun{chakrabarti06evo} proposed a general framework for ``evolutionary clustering'', where both current and historic information was incorporated into the objective function of the clustering process. The authors used this idea to formulate dynamic variants of common agglomerative and partitional clustering algorithms. In the latter case, related clusters were tracked over time by matching similar centroids across time steps. Two evolutionary versions of spectral partitioning for classical (unipartite) graphs were proposed by \citeasnoun{chi07evo}. The first version (PCQ) involved applying spectral clustering to produce a partition that also accurately clusters historic data. The second version (PCM) involved measuring historic quality based on the chi-square distance between current and previous partition memberships. Both algorithms were applied to synthetic data and weekly blog link data. Recently, evolutionary approaches have been extended to the case of co-clustering -- using bipartite spectral partitioning \cite{greene10dynak} and three-way matrix factorization \cite{pen11coclustering}.

\subsubsection{Dynamic Community Finding}
\label{sec:reldyn2}
The application of unsupervised dynamic data exploration methods has been particularly prevalent in the realm of social network analysis, where the goal is to identify groups representing communities of users in dynamic networks. \citeasnoun{palla07quant} proposed an extension of the popular \emph{CFinder} clustering algorithm to identify community-centric evolution events in dynamic graphs, based on an offline strategy. This extension involved applying community detection to composite graphs constructed from pairs of consecutive time step graphs. The resulting clique-based communities are subsequently matched to communities in either of the individual time steps. Another life-cycle model was proposed by \citeasnoun{berger06framework}, where the dynamic community finding approach was formulated as a graph coloring problem. The authors proposed a heuristic solution to this problem, by greedily matching pairs of node sets between time steps, in descending order of similarity. 
\citeasnoun{asur09event} described a matching-based, cluster event identification strategy, which was implemented in the form of bit operations computed on time step community membership matrices. This strategy was applied to both bibliographic networks and clinical trial data in the context of pharmaceuticals. 

\subsection{Visualization}
\label{sec:relviz}
Our visualization approach encodes both cluster and topic evolution, emphasizing
how the collection of clusters evolve over time.  As a 
result, the technique draws on previous work in both text and 
dynamic graph visualization in a novel way to visualize the evolution of
terms and clusters simultaneously.   
In section~\ref {secTextVisPW}, we discuss several ways to visualize static
and dynamic collections of unstructured text documents.  
Section~\ref {secDynamCommPW} discusses
methods for visualizing dynamically evolving communities.

\subsubsection{Static and Dynamic Text Data}
\label {secTextVisPW}
A number of works have looked at the problem of visualizing large collections
of text documents in static and dynamic settings~\cite {09Hearst}.  In this 
section, we cover a subset of these techniques that focus on visualizing
the frequency of a term or theme without context in the collection of 
documents.  Our visualization technique uses this work as a basis for
visualizing word frequency data.

{\bf Static context:} When summarizing word frequencies in a document or 
collection of documents, tag clouds are often used.  Tag clouds have a history
in varied domains including social psychology, literary works, finance, and
web technologies~\cite {08Viegas}.  In a tag cloud, the 
frequency of a term or theme is mapped to the size of the word in the cloud,
and these words are placed on several lines in the display.
Variants of the tag cloud representation exists, 
Wordle\footnote {\url{http://www.wordle.net}}~\cite {09Viegas} for example, where the orientation of
the words is freely chosen and words can be placed inside one another.
As this is the most common method for visualizing term frequency in a document
or collection of documents, we use the standard tag clouds to summarize the
top terms in a cluster of documents and their aggregates.

{\bf Tag cloud evaluation:} A number of user studies have looked at the
effectiveness of tag clouds.  A study~\cite {07Rivadeneira}
presented two experiments that evaluated tag cloud properties such as
font size and word order.  The authors
found that words with larger font sizes were recalled more easily and that
when words were organized according to frequency, participants were better 
able to form an impression about the subjects discussed in the document.  A 
second experiment~\cite {07Halvey} also investigated the time taken to 
find words in a tag cloud.  The experiment 
found that larger tag sizes and an alphabetical ordering of the tags decreased 
task completion time.  Alphabetically ordered lists of words outperformed 
alphabetically ordered tag clouds in terms of completion time for this task.
In TextLuas, we support both frequency and alphabetical orderings.

{\bf Dynamic context:} Often, how a term or a set of terms evolve over time
is the focus of analysis.  In these cases, time is encoded spatially, often
as a flow from left to right, and the frequency of terms are encoded on the
vertical axis.  ThemeRiver~\cite {02Havre} expresses the frequency of a
term or theme over time.  Each theme is given its own 
color and stream thickness encodes frequency.  As time flows from left to right,
the changes in thickness encode the changes in frequency.  The 
NameVoyager~\cite {05Wattenberg} was designed in a similar way.  In the tool,
time flows left to right and the thickness of streams indicate the 
frequency of names given to children born that year.  However, neither system
encodes dynamic cluster structure and instead focuses on term/name evolution
only.

\subsubsection{Dynamic Community Visualization}
\label {secDynamCommPW}
As noted in \refsec{sec:reldyn2}, a common application of dynamic clustering methods has been in the task of finding communities in social networks. When the number of communities and the size of each community is small or
medium-sized,
visualization techniques exist to examine how the topologies of those
communities evolve over time~\cite {04Frishman} or diagrams that 
depict the evolution of community structures~\cite{tanti07framework}.  
However, neither method directly handles terms associated with each community, and the representations can have difficulty scaling data sets with larger
community sizes and a larger number of communities.

More abstract representations of communities and their structure are often 
needed to allow visualizations to scale to a larger number of communities.
\citeasnoun{06Falkowski} creates a node for each community.  The x-coordinate of
that node corresponds to its time period.  The y-coordinate is computed based 
on similarity to other communities at that time and directed edges connect
communities in adjacent time periods to describe community evolution.
Alluvial diagrams~\cite{rosvall10change} depict each community as a
line swath whose width is proportional to its number of members.  Gaps 
are placed between time periods and the swaths split and merge as the
communities evolve over time.  On the other hand, \citeasnoun{08Yang} present
a visual analytics system that is more topology-centric.  An overview of 
all major events in community evolution is presented, and the user of the 
system can click on elements of this list to
view what happened to the community before and after the event.

This previous work either employs a 
topology-centric~\cite {04Frishman,08Yang} or 
an evolution-centric~\cite{06Falkowski,tanti07framework,rosvall10change} 
approach for depicting community evolution, by using
spatial position to encode either community topology or evolution
respectively.  In our approach, we start with a 
evolution-centric view of the data.  However, because we only deal with
clusters of documents and terms, we are able to provide
summaries, in terms of textual content of the clusters, through tag clouds.

\section{Methods}
\label{sec:methods}

\subsection{Overview}
\label{sec:prob}

In the offline formulation of the dynamic co-clustering problem, our overall goal is to identify a set of \emph{dynamic clusters}, represented as linear timelines extending across multiple time steps. \emph{Step clusters} are the clusters identified at individual time steps. These represent specific observations of dynamic clusters at a single point in time.
The offline dynamic text co-clustering problem has three key requirements: 
\begin{enumerate}
\item An approach to aggregate step clusters into dynamic cluster timelines, and track these dynamic clusters across time steps. Based on our previous work in social network analysis \cite{greene10tracking}, we propose a framework for tracking clusters in dynamic text data in \refsec{sec:model}, and describe a specific implementation of the framework in \refsec{sec:track}.
\item A suitable algorithm to cluster individual time step graphs. We briefly describe a suitable choice of algorithm in \refsec{sec:cocluster}. 
\item An appropriate technique to visualize dynamic cluster timelines, highlighting both events in the evolution of clusters and representative cluster content. In \refsec{sec:tagging} we describe the problem of generating content summaries for clusters.  The visualization techniques are discussed in \refsec{sec:viz}.
\end{enumerate}

\subsection{Tracking Dynamic Clusters}
\label{sec:model}

\subsubsection{Problem Formulation}
As described in \refsec{sec:reldoc}, previous work by \citeasnoun{dhillon01cocluster} involved representing a static text data set in the form of a single weighted bipartite graph. Documents and terms can then be co-clustered by partitioning the nodes of the bipartite graph.
For the case of dynamic text data, where we are following an offline strategy, a complete corpus can be represented as a set of $l$ bipartite graphs $\fullset{G}{l}$, where each graph is built from the documents collected during a single time step. The duration of the time step controls the ``resolution'' at which we intend to explore the data. Each \emph{step graph} $G_{t}$ consists of two distinct sets of nodes, representing the $m_{t}$ terms and $n_{t}$ documents present in the data at time $t$. Clustering each step graph yields a set of \emph{step clusters}.  These clusters contain both terms and documents, and may be disjoint or overlapping. The dynamic clustering task involves identifying a set of \emph{dynamic clusters} built from step clusters observed at different time steps. 

\begin{figure}
\begin{center}
\includegraphics[width=0.8\linewidth]{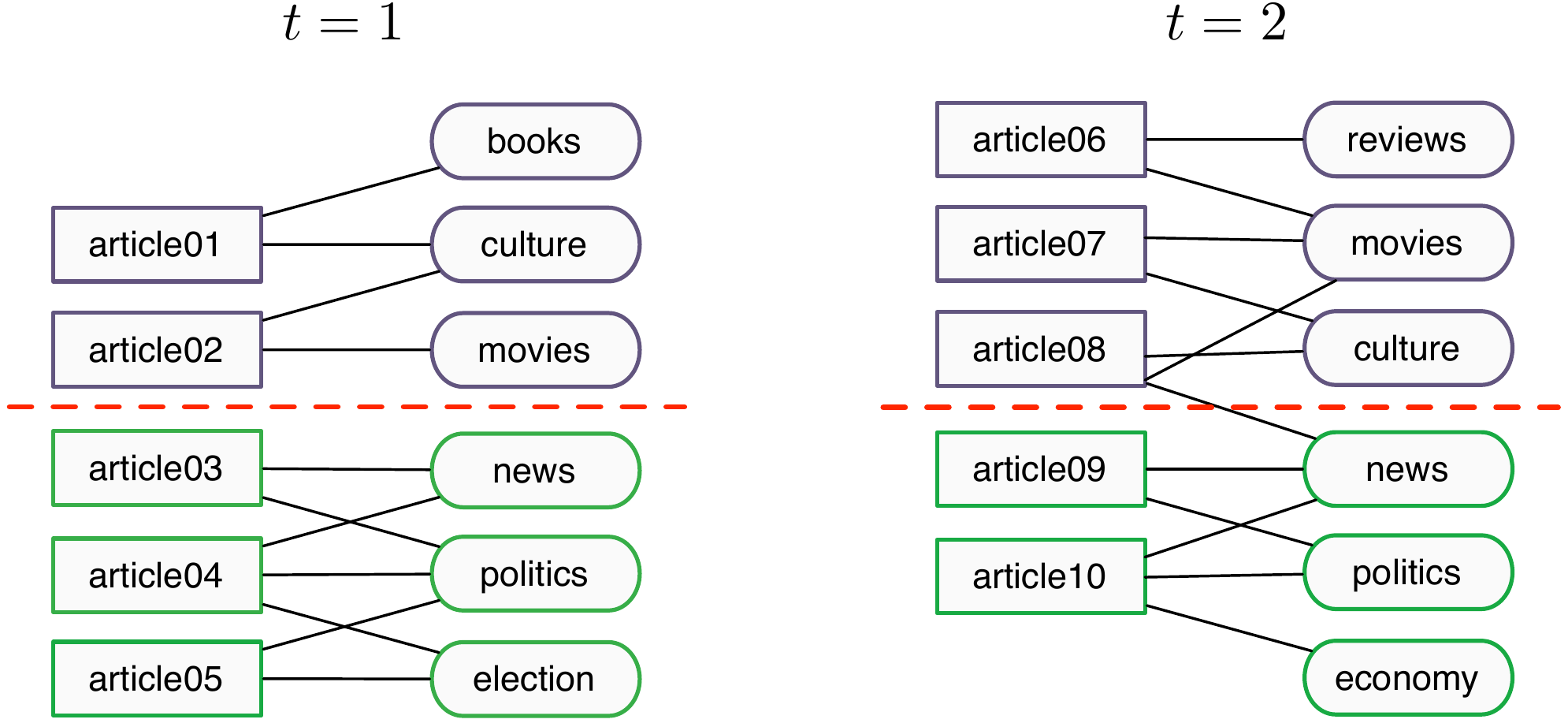}
\caption{\label{fig:bipartite}Example of a dynamic text data set, which consists of two time steps, ten documents, and eight unique terms. The data set is represented in the form of two bipartite graphs, showing the presence of terms in documents in both time steps. Note that a subset of terms alone persist between the two time steps. A trivial partitioning of both documents and terms across the two time steps is indicated by the dashed line.}
\end{center}
\end{figure}

In some domains, both data objects and features will persist over time. However, for text-based data it will generally be the case that documents will only exist at given point in time (\eg the publication date for news articles, research papers), while many terms will persist in the data over time. Therefore, while step clusters contain both terms {\textbf{and}} documents, we propose tracking dynamic clusters (\ie connecting step clusters) based on terms alone.  These terms represent indicative keywords that define a particular theme, topic, or news story across time.

An illustrative example is shown in \reffig{fig:bipartite}, where a dynamic text data set is represented by two bipartite graphs corresponding to two discrete time steps. In each bipartite graph there is a trivial partitioning of both documents and terms into two clusters. By matching clusters via the subset of terms common to the two graphs, we can readily see that the two topical clusters persist across both time steps.  These represent dynamic clusters in the data.

\subsubsection{Dynamic Cluster Timelines}
\label{sec:timeline}
We would like to define a convenient structure for representing the history and development of dynamic clusters in a data set. Firstly, we denote the set of $k'$ dynamic clusters as $\metaset{D}=\fullset{D}{{k'}}$, and denote the set of $k_{t}$ step clusters identified at time $t$ as $\metaset{C}_{t}={\{C_{t1},\dots,C_{tk_t}\}}$.
Consequently, each dynamic cluster $D_{i}$ can be represented by a \emph{timeline} of its constituent step clusters, ordered by time, with at most one step cluster for each time $t$. 

\begin{figure}
\begin{center}
\includegraphics[width=0.665\linewidth]{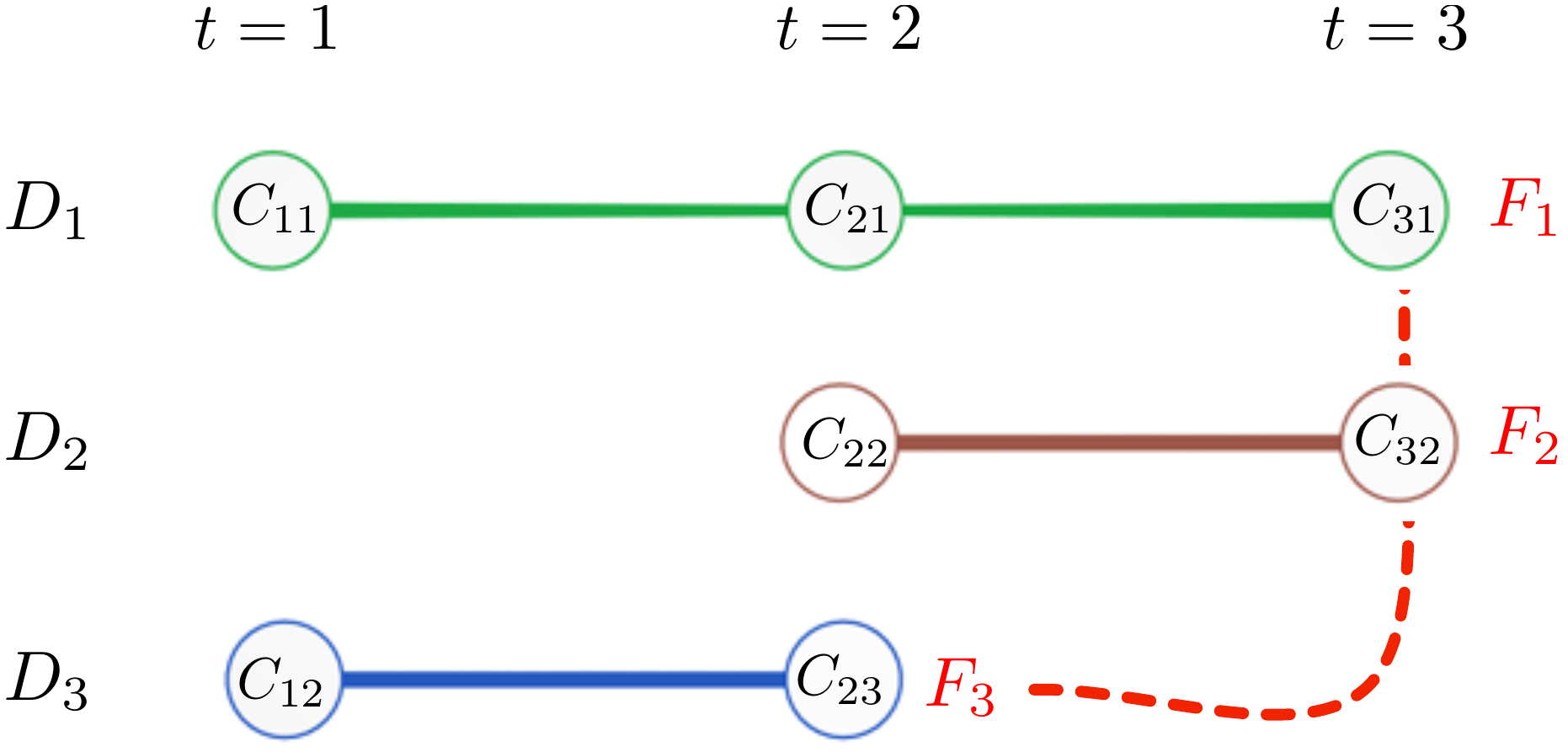}
\caption{\label{fig:example1}Example of three dynamic cluster timelines tracked over three time steps, featuring continuation, birth, and death cluster life-cycle events. The dynamic cluster \emph{fronts} are highlighted. These represent the most recent observations of the dynamic clusters.}
\end{center}
\end{figure}

\begin{figure}
\begin{center}
\includegraphics[width=0.47\linewidth]{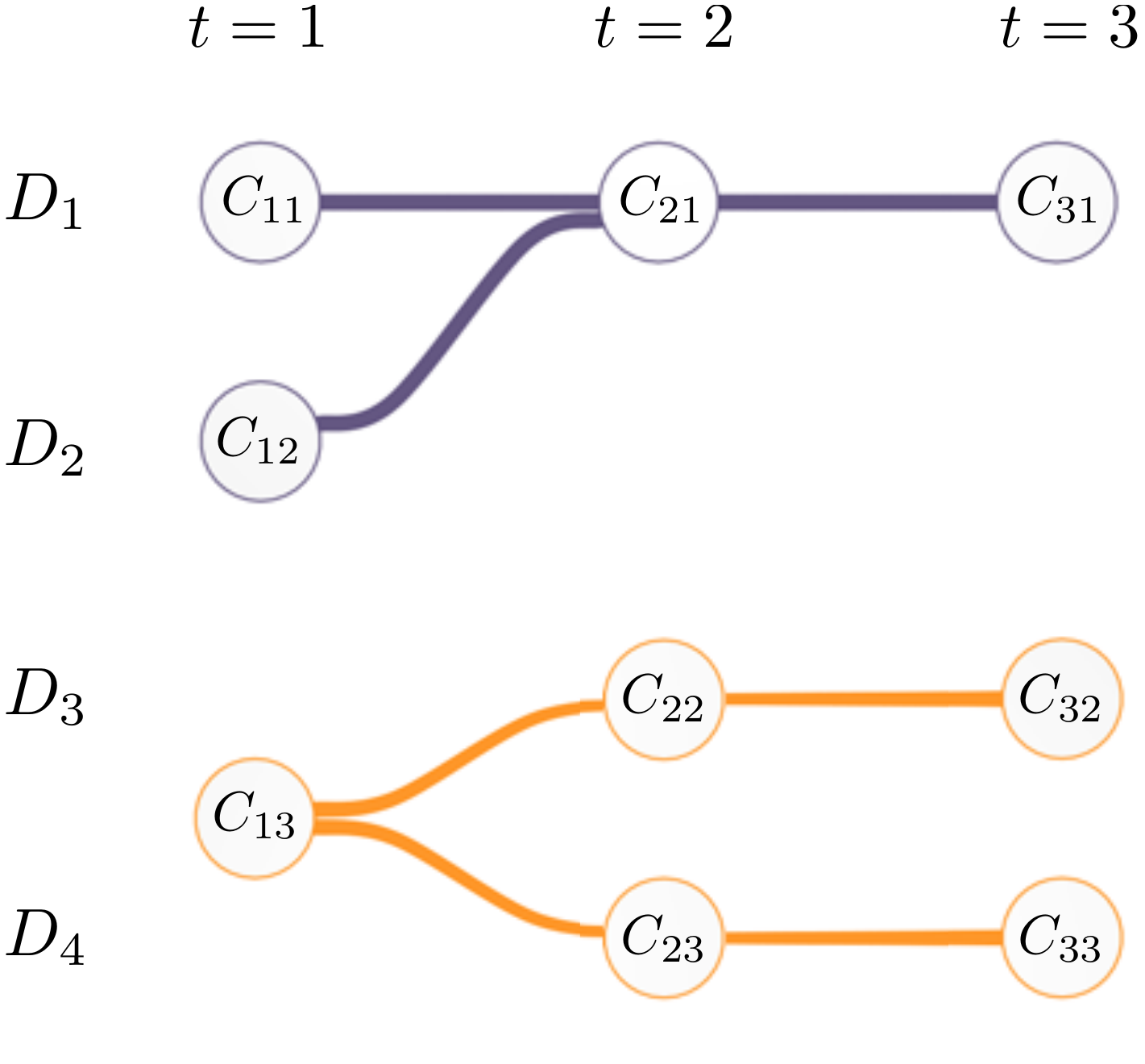}
\caption{\label{fig:example2}Example of four dynamic cluster timelines over three time steps, illustrating merge and split life-cycle events.}
\end{center}
\end{figure}

\reffig{fig:example1} shows a simple example involving three step clusterings containing observations of three distinct dynamic clusters. The timelines for these three dynamic clusters are straight-forward:
\begin{itemize}
\item $D_{1}$: \{$C_{11},C_{21},C_{31}$\}
\item $D_{2}$: \{$C_{22},C_{32}$\}
\item $D_{3}$: \{$C_{12},C_{23}$\}
\end{itemize}
A more complex scenario is shown in \reffig{fig:example2}. Note that, while there appear to be three distinct branches at time $t=3$, there are in fact four dynamic clusters with four corresponding timelines as follows:
\begin{itemize}
\item $D_{1}$: \{$C_{11},C_{21},C_{31}$\}
\item $D_{2}$: \{$C_{12},C_{21},C_{31}$\}
\item $D_{3}$: \{$C_{13},C_{22},C_{32}$\}
\item $D_{4}$: \{$C_{13},C_{23},C_{33}$\}
\end{itemize}
The most recent observation in a timeline is referred to as the \emph{front} of the dynamic cluster.  The front for $D_{i}$ is denoted $F_{i}$. The fronts for the three dynamic clusters are highlighted in \reffig{fig:example1}.  Note that the dynamic cluster $D_{3}$ does not have a corresponding observation at time $t=3$, so its front is the step cluster $C_{23}$ from the previous time $t=2$.


\subsubsection{Dynamic Cluster Events}
\begin{figure}
\begin{center}
\includegraphics[width=4.65in]{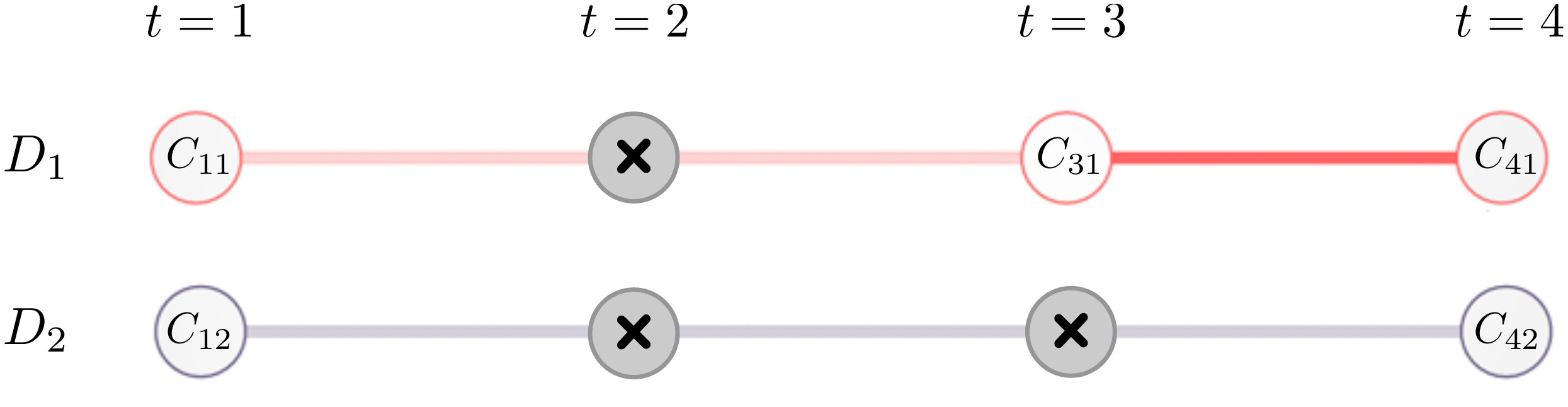}
\caption{\label{fig:example3}Example of two ``intermittent'' dynamic clusters which are not observed at every time step after birth. The dynamic cluster $D_{1}$ is unobserved in the graph at time $t=2$, but continues in time step $t=3$, while the dynamic cluster $D_{2}$ is missing from both $t=2$ and $t=3$. }
\end{center}
\end{figure}

When exploring the themes or topics present in a document collection, we would like to examine their development over time and the relationships between them. Based on the timeline structures, we can characterize the evolution of dynamic clusters in terms of a number of fundamental cluster ``life-cycle'' events:
\begin{itemize}
\item \emph{Birth:} The emergence of a step cluster $C_{tj}$ observed at time $t$ for which there is no corresponding dynamic cluster in $\metaset{D}$. A new dynamic cluster $D_{i}$ containing $C_{tj}$ is created and added to $\metaset{D}$. An example in \reffig{fig:example1} is the cluster $D_{2}$ born in the second time step.
\item \emph{Death:} The dissolution of a dynamic cluster $D_{i}$ occurs when it has not been observed (\ie there has been no corresponding step cluster) for at least $d$ consecutive time steps. $D_{i}$ is subsequently removed from the set $\metaset{D}$. An example in \reffig{fig:example1} is $D_{3}$, assuming that no further step clusters are subsequently assigned to its timeline.
\item \emph{Merging:}  A merge occurs if two distinct dynamic clusters $(D_{i},D_{j})$ observed at time $t-1$ match to a single step cluster $C_{ta}$ at time $t$. The pair subsequently share a common timeline starting from $C_{ta}$.  In \reffig{fig:example2} the dynamic clusters $D_{1}$ and $D_{2}$ are both matched to $C_{21}$ in the second step.
\item \emph{Splitting:}  It may occur that a single dynamic cluster $D_{i}$ present at time $t-1$ is matched to two distinct step clusters $(C_{ta},C_{tb})$ at time $t$. A branching occurs with the creation of an additional dynamic cluster $D_{j}$ that shares the timeline of $D_{i}$ up to time $t-1$, but has a distinct timeline from time $t$ onwards. In \reffig{fig:example2} an existing dynamic cluster $D_{3}$ is matched to both $C_{22}$ and $C_{23}$ in the second step, resulting in the creation of an additional dynamic cluster $D_{4}$.
\end{itemize}
In most cases, we will observe trivial \emph{continuation} events, where a dynamic cluster observed at time $t$ also has an observation at time $t+1$. However, note that a dynamic cluster may not necessarily be observed at all time steps after birth -- it may be observed at birth time $t$ and at death time $t' > t$, but may be missing from one or more intermediate steps (but less than $d$ consecutive steps) between $t$ and $t'$. Two examples are shown in \reffig{fig:example3}. This reflects the idea that temporally ``intermittent'' clusters can exist in the data. For instance, in the case of mainstream news data, a story may receive considerable coverage for a short period of time, coverage may then cease and subsequently recommence at a later date when there are further developments relating to the story. The frequency of appearance of such intermittent behavior will often depend on the duration of each time step.

\subsubsection{Tracking Clusters Across Time Steps}
\label{sec:track}

The model presented above describes a general framework for representing dynamic clusters, in the form of timelines of step clusters that highlight cluster life-cycle events. Given a set of bipartite graphs representing a complete dynamic text data set, we now propose a simple approach for constructing such timelines. The key issue in this task concerns how best to match step clusters at a given time $t$ to the existing set of dynamic clusters $\metaset{D}$. 
To address this issue, we employ a heuristic threshold-based matching method, which allows for many-to-many mappings between clusters across different time steps. This mapping supports the identification of dynamic events such as cluster merging and splitting.
As noted previously, although we cluster terms and documents at each step, we intend to track dynamic clusters based on terms persisting across different time steps. Therefore, the mappings are calculated solely based on the term-cluster memberships in step clusterings. 

The proposed approach proceeds as follows. The first step clustering $\metaset{C}_1$ of terms and documents is generated by applying a suitable clustering algorithm (such as that described next in \refsec{sec:cocluster}) to the bipartite graph $G_{1}$.  We use this graph to bootstrap the process. A distinct dynamic cluster is created from the term memberships for each step cluster. The next step clustering $\metaset{C}_2$ is generated by clustering the bipartite graph $G_{2}$. We then attempt to match these step clusters with the dynamic cluster fronts $\fullset{F}{{k'}}$ (\ie the step clusters from $\metaset{C}_1$). All pairs $(C_{2a},F_{i})$ are compared, and the dynamic cluster timelines and fronts are updated based on the event rules described previously in \refsec{sec:model}. The process continues until all $l$ step graphs have been processed.

To perform the actual matching, we employ the widely-used Jaccard coefficient for binary sets \cite{jaccard12index}. Given a step cluster $C_{ta}$ and a dynamic cluster front $F_{i}$, the similarity between the pair is calculated as:
\begin{equation} 
sim(C_{ta},D_{i}) = \frac{\vert T_{ta} \cap F_{i} \vert}{\vert T_{ta} \cup F_{i} \vert} 
\label{eqn:jac}
\end{equation}
where $T_{ta}$ denotes the set of terms in the step cluster $C_{ta}$.
If the similarity exceeds a matching threshold $\theta \in [0,1]$, the pair are matched, and $C_{ta}$ is added to the timeline for the dynamic cluster $D_{i}$ corresponding to $F_{i}$. 

The output of the matching process between a step clustering $\metaset{C}_{t}$ and the existing fronts $\fullset{F}{{k'}}$ will implicitly reveal a series of cluster evolution events. A step cluster $C_{ta}$ matching to a single dynamic cluster indicates a continuation, while the case where $C_{ta}$ matches multiple dynamic clusters results in a merge event. If no suitable match is found for $C_{ta}$ above the threshold $\theta$, a new dynamic cluster is created for $C_{ta}$. 

\subsection{Clustering at Individual Time Steps}
\label{sec:cocluster}

Previously we have described an approach for constructing dynamic cluster timelines from collections of step clusters. We also require a suitable clustering algorithm to produce a clustering of each step graph. One option is to apply an existing clustering or community-finding algorithm on a unipartite representation of the data, and then derive term memberships from these groups and the original data (\eg by examining term-centroid weights). An alternative is to apply a co-clustering approach to cluster both documents and terms simultaneously \cite{dhillon01cocluster}. However, previous work has shown that there are advantages to clustering techniques that also consider historic information, rather than taking a static approach which treats each time step entirely in isolation \cite{chakrabarti06evo}.

We have previously introduced a dynamic spectral co-clustering algorithm \cite{greene10dynak}. This algorithm takes into account both information from a spectral embedding of the current bipartite graph, together with historic information from the previous time step. Evaluations performed on news and social media data sets showed that the algorithm is effective in accurately identifying coherent clusters, while also ensuring a consistent transition between clusterings in successive time steps. In this section, we provide a general overview of the algorithm, and apply it as part of our system evaluation in sections \ref{sec:evalreal1} and \ref{sec:evalreal2}.

%

Firstly, we represent the bipartite graph $G_{t}$ for step $t$ as a rectangular adjacency matrix $\m{A}_{t}$ of size $m_{t} \times n_{t}$, with rows corresponding to terms and columns corresponding to documents. We construct the degree-normalized adjacency matrix $\m{\hat{A}}_t = \m{D_1}^{-\frac{1}{2}} \m{A}_{t} \m{D_2}^{-\frac{1}{2}}$, where $\m{D_1}$ and $\m{D_2}$ are diagonal column and row degree matrices respectively. We then apply SVD to $\m{\hat{A}}_t$, computing the leading left and right singular vectors corresponding to the largest singular values. We use $k_{t}$ singular vectors, corresponding to the expected number of clusters for the time step. The truncated SVD yields matrices $\m{U_{k_{t}}}$ and $\m{V_{k_{t}}}$. A unified embedding of terms and documents is constructed by normalizing and stacking the truncated factors as follows:
\begin{equation}
\m{Z}_{t} = \left[ 
\begin{array}{ccc}
\m{D_1}^{-1/2} \m{U_{k_{t}}} \\
\m{D_2}^{-1/2} \m{V_{k_{t}}} 
\end{array} \right]
\label{eqn:stack}
\end{equation}
Prior to clustering, the rows of $\m{Z}_{t}$ are subsequently re-normalized to have unit length, as proposed by \citeasnoun{ng01spectral}. The rows of this matrix provide a $k_{t}$-dimensional embedding of all terms and documents present in $G_{t}$.

At time $t=1$, we have no historic information. So we generate a step clustering by applying $k$-means to the embedding $\m{Z}_{1}$, using orthogonal initialization \cite{ng01spectral}. For $t > 1$, we wish to initialize using clusters from the previous time step. Since we assume that only a subset of terms will persist between time steps, we will not have initial cluster memberships for any documents, and may also lack memberships for some terms that were not present in the last step. Therefore, we predict initial cluster memberships for each unassigned row $z_{i}$ of $\m{Z}_{t}$, using a simple nearest centroid classifier trained on the centroids constructed from rows of $\m{Z}_{t}$ for which membership information is available. This yields a predicted clustering $\aset{P}_{t}$ based on historic information.

Once we have initialized the clustering process with $\aset{P}_{t}$, we apply a constrained version of $k$-means clustering to the rows of the embedding $\m{Z}_{t}$, taking into account both the internal quality of the current partition and agreement with the predicted partition $\aset{P}_{t}$.
To combine both sources of information, the clustering objective becomes a weighted combination of two quality measures:
\begin{equation}
J(\aset{C}_{t}) = (1-\alpha) \cdot  \left( \sum_{c=1}^{k} \sum_{z_i \in C_c} \vectrans{z_i} \vec{\mu_c} \right )
+ \alpha \cdot  \left( \textrm{pred}(\aset{P}_{t},\aset{C}_{t}) \right)
\label{eqn:obj}
\end{equation}
The first term in \reft{eqn:obj} represents the standard spherical $k$-means objective \cite{dhillon01sphere}, while in the second term, $\textrm{pred}(\aset{P}_{t},\aset{C}_{t})$ denotes the pairwise agreement  between the predicted clustering $\aset{P}_{t}$ and the current clustering. The parameter $\alpha \in [0,1]$ controls the balance between the influence of the information present in the current spectral embedding and the historical information. A higher value of $\alpha$ allows information from the previous time step to have a greater influence. The output of the constrained $k$-means process is a co-clustering of the terms and documents in $G_{t}$.

\subsection{Post-Processing for Visualization}
\label{sec:tagging}

\subsubsection{Building Tag Clouds}
As described in \refsec{sec:relviz}, one way to summarize a collection of text data is to use a tag cloud. In this paper, we use tag clouds to present an overview of the content in a single step cluster or the aggregate content of multiple step clusters (\eg the step clusters from one or more dynamic clusters.).  The visualization technique will be further described in \refsec{sec:viz}.

In its most fundamental form, the data represented in a tag cloud consists of a set of unique terms (\ie the tags) with a corresponding set of weights, indicating the relative importance of each term. These weights are used to control the prominence of each term in the cloud. To identify the set of descriptive terms for each step cluster, we use a variant of the centroid-based ``concept decomposition'' method proposed by \citeasnoun{dhillon01sphere}.  For a step cluster $C_{ta}$ containing terms and documents, we 
calculate the term weights as follows:
\begin{enumerate}
\item Based on the documents assigned to $C_{ta}$, compute the centroid vector $w_{ta}$ from the term-document frequency values in the corresponding step graph $G_{t}$.
\item Set all entries in $w_{ta}$ to zero, with the exception of those entries corresponding to the terms assigned to $C_{ta}$.
\item The terms with the highest non-zero weights in $w_{ta}$ are used to produce the tag cloud representing the content in $C_{ta}$.
\end{enumerate}

\subsubsection{Aggregating Tag Clouds}
To generate a tag cloud describing the aggregate content for two or more step clusters, we construct a weight vector for each cluster as described above and compute the mean vector $\bar{w}$. The highest-weighted terms in $\bar{w}$ provide the tags for the aggregate cloud. In this way, we can summarize the content of part or all of a dynamic cluster timeline.

\section{Visualization}
\label{sec:viz}
The model and the output of the dynamic clustering approaches, discussed in 
section~\ref {sec:model}, naturally leads toward visualization techniques
which explicitly encode timelines.  Visualizations, structured in this way, 
would provide a life-cycle perspective of the data.  In our case, 
we employ a ``metro map'' metaphor to depict dynamic cluster evolution.  

\begin {figure}[!b]
\centering
\includegraphics[width=0.85\linewidth]{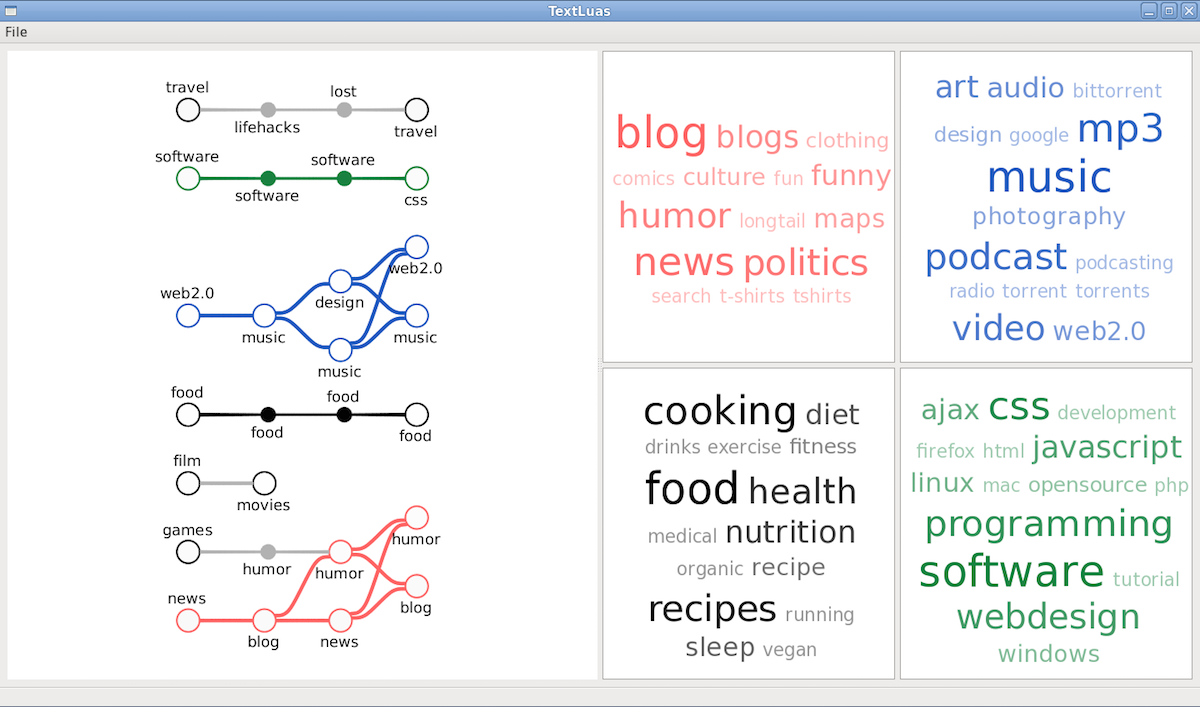}
\caption {TextLuas main interface.  The left side of the screen is the
metro map view which depicts the dynamic clusters.  The right side of the
screen is the matrix of summary tag clouds which summarizes the terms
present in each line.}
\label {figVisMainScreen}
\end {figure}

The visualization is composed of three components.  The 
first two components, presented in Fig.~\ref {figVisMainScreen}, provide 
overviews of cluster evolution.  The set of cluster timelines on the left, depicting 
the events in the evolution of timelines, is the \emph{metro map view}, which is
discussed in section~\ref {secMetroMap}.  On the right of this same display
is the \emph{matrix of summary tag clouds}.  These summaries, discussed in 
section~\ref {secSummaryTagClouds}, have the same color of the line that 
they represent and have a tag cloud that encodes the average frequency of
the terms on that line.  Clicking on any tag cloud in this matrix, brings up 
the \emph{metro line details}.  This view, shown in 
Fig.~\ref {figDetailsView}, provides details for a single timeline or the
parts of the metro map that are in the same color.  This view is discussed
in section~\ref {secMetroDetail}.

Our system heavily relies on linked views~\cite {00North}.  Linked views
have the advantage of maximizing the readability of the summary tag clouds
and the textual content in the details view.  Integrating the tag clouds
into a single visualization would have the advantage that only a single
view would be required for visualizing the data.  However, they would have
the disadvantage that reading the textual content of a large number of
timelines would be difficult.

\subsection {Metro Map View}
\label {secMetroMap}
The metro map view shows the dynamic cluster timelines for a complete dynamic text data set, through the use of a metro map
metaphor.  In this view, time progresses from left to right, with $t=1$
being in the leftmost part of the diagram.  
Initially, all lines are colored gray.
The user of the system can select metro lines to change their color.  
Summaries of these lines are added to the summary tag cloud matrix.  The
next two sections describe how this diagram is drawn, and how users interact
with it.

\subsubsection {Diagram Rendering and Encoding}
In our metaphor, step clusters are indicated by ``stations''.  A directed edge connects two step clusters if they appear successively in the same timeline. Thus, the 
topology of any dynamic cluster timeline is always a 
directed acyclic graph (DAG), as edges are always oriented in the direction 
of increasing time (left to right in our diagrams).  The problem of drawing 
DAGs is well-studied, with many algorithms existing to
generate useful drawings.  In this work, we use the algorithm implemented
in the Tulip framework~\cite {gdauber} as a basis for diagram generation.
All metro map style drawings in this paper are also rendered using Tulip.

We would like to ensure that all step clusters
co-occurring at the same time $t$ are vertically aligned.  As dynamic 
clusters can be born, become intermittent, and die at any time, for the
purposes of diagram layout we need to
insert dummy nodes.  For birth events that happen at times $t > 1$,
we prefix the birth event with a series of dummy nodes.  For intermittent
events, we connect the last time the cluster was observed to the next time
the cluster is observed with a series of dummy nodes (the nodes marked ``x''
in Fig.~\ref {fig:example3}).  For deaths, we do not insert dummy nodes 
after the last observation of the cluster and the algorithm places the 
node at the correct level.  Dummy nodes are inserted for the purposes of
diagram layout, but these nodes are removed before diagram display.

When an edge connects two nodes that are not horizontally aligned, it is
represented as a smoothly interpolating spline curve.  The line-projected
control points of this spline curve are placed procedurally according to the 
distance between the stations.  For stations $(x_1, y_1)$ and $(x_2, y_2)$ with
the vector between them $(a, b)$, the two 
control points are placed at $(x_1 + a/3, y_1 + b/6)$ and 
$(x_2 - a/3, y_2 - b/6)$.  If the degrees of the nodes in the diagram are
relatively low, this heuristic allows edges smoothly bend into and out of 
stations.  

\begin {figure}
\centering
\subfigure [Major] {\includegraphics[width=0.18\linewidth]{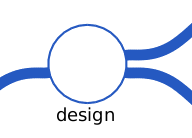}\label {majorStationEncoding}}
\subfigure [Minor] {\includegraphics[width=0.18\linewidth]{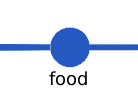}\label {minorStationEncoding}}
\subfigure [Intermittent] {\includegraphics[width=0.24\linewidth]{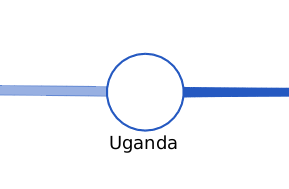}\label {intermittantLine}}
\caption {Encoding for major stations, minor stations, and intermittent lines.
{\bf (a)} Major station is depicted as a hollow circle.  {\bf (b)} Minor 
station is depicted as a solid circle.  {\bf (c)} Intermittent lines are have
a lower alpha value than lines which represent continuation.  In this 
subfigure, the left line is intermittent while the right line is not.}
\end {figure}

In the diagram, there are major stations and minor stations.  A 
\emph{major station} corresponds to a step cluster at which a life-cycle event
occurs:  birth, death, merge, split, or begin/end of an intermittent
period.  Major stations are drawn as hollow circles as shown in 
Fig.~\ref {majorStationEncoding}.  \emph{Minor stations} correspond to 
step clusters that are continuation events.  They always
have an in degree of one and an out degree of one and are not adjacent to
any dummy nodes.  These stations are
encoded as filled circles as shown in Fig.~\ref {minorStationEncoding}.
When a dynamic cluster becomes intermittent, its edge is rendered with
a higher alpha value as shown in Fig.~\ref {intermittantLine}.
Stations, by default, are named with the term that is the most prevalent in 
its cluster.  However, the user of the system can specify a ``names'' file
as input for custom station names that would be more useful for their task.
For split and merge events, this label is always placed beneath the node.  
For all other stations, the label placement alternates between above and
below the station in the depiction of the dynamic cluster.

\subsubsection {Interaction}
\begin {figure}
\centering
\subfigure [Before Selection] {\includegraphics[width=0.30\linewidth]{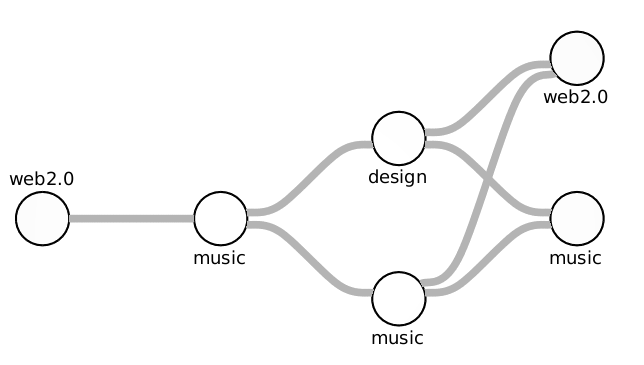}\label {beforeSelection}}
\subfigure [After Selection] {\includegraphics[width=0.30\linewidth]{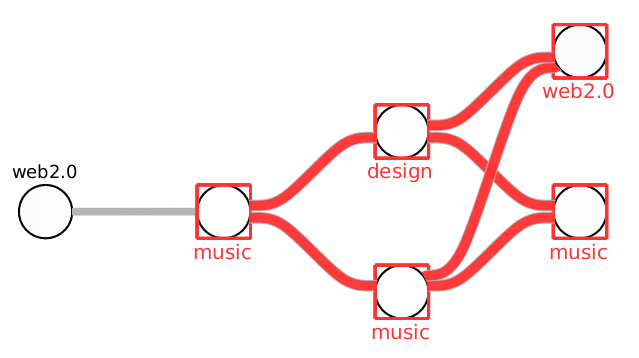}\label {afterSelection}}
\subfigure [After Deselection] {\includegraphics[width=0.30\linewidth]{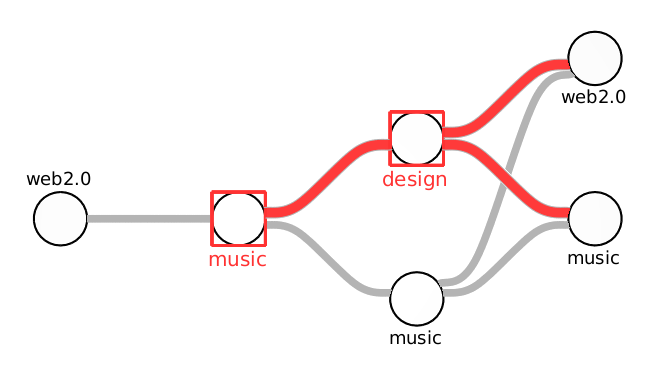}\label {afterDeSelection}}
\caption {Figure showing flood selection in TextLuas.  {\bf (a)} Unselected
portion of the metro map. {\bf (b)} After clicking on ``music'' at time
2, all nodes rightward and connected to ``music'' are selected.  {\bf (c)}
After clicking on the edge between ``music'' $t = 2$ and ``music'' 
$t = 3$, the lower subtree is deselected.  The leaves, ``music'' and 
``web2.0'', can be re-added to the selection by clicking on them individually.}
\end {figure}

As selecting all stations individually can be time consuming, selection
and deselection of stations and edges in the TextLuas is propagated rightward 
following the direction of the edges.  Fig.~\ref {beforeSelection} shows 
part of the metro map.  By clicking on the node labeled ``music'' at $t = 2$,
the entire subtree connected to the right of ``music'' is selected as 
shown in Fig.~\ref {afterSelection}.  Subtrees can be removed from the
selection in a similar way.  Fig.~\ref {afterDeSelection} shows the result
after the edge between ``music'' ($t = 2$) and ``music'' ($t = 3$) was clicked,
removing that subtree from the selection.  If the user would like to re-add
the nodes ``music'' and ``web2.0'' at $t = 4$, they can be selected manually.
Once the user is happy with their selection, they simply press a number key
to color the entire selection. It is important to note that selections do
not need to be connected.

Initially, all lines in the diagram are colored gray.  When a metro line takes
on a color, its corresponding tag cloud is placed in the summary of tag
clouds matrix with all nodes colored the same color as the metro line.

\subsection {Matrix of Summary Tag Clouds}
\label {secSummaryTagClouds}
The matrix of summary tag clouds presents an overview of the topics 
discussed on each colored timeline of the TextLuas.  Tag size is based on the 
term weighting strategy described in 
section~\ref {sec:tagging},
using the mean of the weight vectors for all step clusters assigned that
color.  The terms of the tag clouds can be ordered
either alphabetically or by frequency to better support a wider range
of tasks.  To see the details of a line, the user simply clicks on the tag 
cloud to bring up the metro line details view, which is discussed in the next 
section.

\subsection {Metro Line Details}
\label{secMetroDetail}
The metro line details view provides a more comprehensive summary of the content and evolution of the metro line (\ie dynamic cluster timeline) of a particular color.
In this view, the user can explore the details of paths on the line, and
the details of the terms at each station.  Fig.~\ref {figDetailsView}
shows the interface, where the top of the screen is the path view
and the bottom is the station cloud matrix.

\subsubsection {Path View}
The path view shows all stations of a particular color and how 
they interconnect with one another.  When the view is launched, a single path is 
highlighted in the display.  Using the up and down arrow keys, the user can
browse through the paths in the diagram in a depth first traversal
order.  For each station highlighted in this view, information is displayed
in the station cloud matrix.
In the top of Fig.~\ref {figDetailsView}, the blue line is shown.  Currently
the path highlighted in $t \in [1,4]$ is ``web 2.0'', ``music'',
``design'', and ``music''.

\subsubsection {Station Cloud Matrix}
\begin{figure}
\centering
\includegraphics[width=0.85\linewidth]{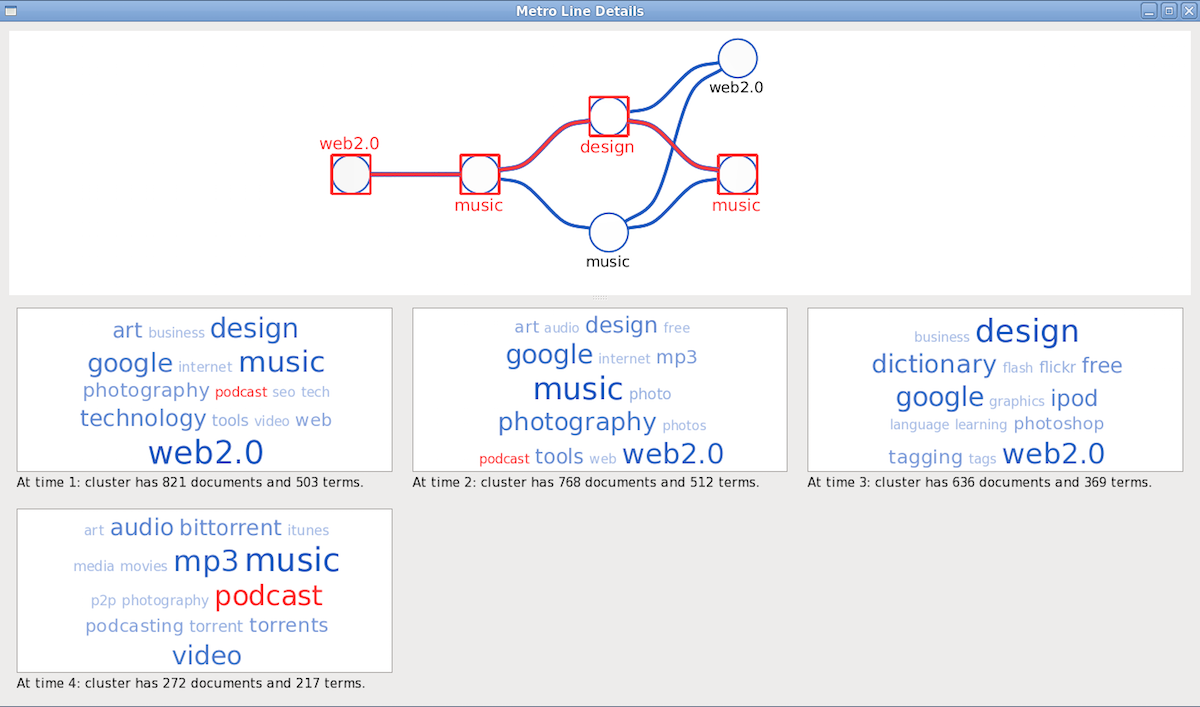}
\caption {Metro line details view.  The display shows the metro line of
the selected color and highlights a path through it in the top half of the
display.  In the bottom half of the display, a small multiples view of
the data at each station along the highlighted path is shown.  Linked 
highlighting between the station clouds indicate the evolution of importance
of a term.  In this example, ``podcast'' is highlighted in all stations except
station three where it falls out of prominence.}
\label {figDetailsView}
\end {figure}

The station cloud matrix is a small multiples~\cite {90Tufte} view --
a matrix of images shows the differences between objects at the given
times, depicting dynamic evolution along that path.  The user of the system
can click on any tag in any of the station clouds to reveal how that tag
evolves in prominence through time.  In Fig.~\ref {figDetailsView}, the
term ``podcast'' has been double clicked, revealing how it evolves down this
line.  Notice that ``podcast'' is not present in the third time step, and this fact is made immediately apparent using the pre-attentive nature of color.  
Recent experiments have shown that for multi-dimensional data (probably the closest type of data to the information we want to visualize per cluster), small multiples seems to be better in terms of human performance \cite{08Robertson}.  
Therefore,
we chose this representation over an animated tag cloud.

In the current tag cloud view, details for the highlighted path are
shown.  The number of documents and terms for the current step
cluster are printed at the bottom of the tag cloud display.  The largest
term in the tag cloud corresponds to the station name, but for the remaining
tags, the small multiples view displays how the term frequency varies along 
the path.
For example, we notice that ``photography'' is a popular term at $t \in [1,2]$,
but becomes less prominent in $t \in [3,4]$.  Also, the term ``video''
suddenly emerges at $t = 4$, when it was not present at any other time. 
At most six station tag clouds are displayed at a time.  The left and right
arrow keys return/advance to the previous/next stations on the line.

\section{Results and Discussion}
\label{sec:eval}
In this section, we present use cases for the TextLuas system in terms of its capacity to visualize dynamically evolving clusters using the techniques described in \refsec{sec:viz}. The experiments are performed on two real-world corpora containing time-stamped documents. The corpora and system implementations are available online\footnote{\url{http://mlg.ucd.ie/textluas}}. Subsequently, in \refsec{sec:UserFeedback}, we provide user feedback based on an application of TextLuas to the exploration of the evolution of scientific communities via the analysis of dynamic bibliographic data.

\subsection{Social Bookmarking Data}
\label{sec:evalreal1}

In our first case study, we considered a Web 2.0 data exploration problem, where a collection of bookmarked websites has been manually assigned terms or tags by a community of users over time. For the purpose of clustering, each site can be described using a ``bag of tags'' text representation. The goal is to cluster sites and terms to explore the changing nature of user interests. Note that, while some bookmarked sites may persist over time, we focus here on tracking dynamic clusters based on terms alone, as described in \refsec{sec:model}. 

We use a subset of the most recent data from a bookmark collection harvested by \citeasnoun{gorlitz08pints} from the \emph{Del.icio.us} web portal. The subset covers the 2,000 top tags and 5,000 top sites across an eleven month period from January to November 2006. We divided this period into 44 weekly time steps, and, for each time step, we constructed a bipartite graph. The two node types correspond to terms (tags) and sites, and edges denote the number of times each site was assigned a given term during the time step. On average, each graph contained approximately 3750 sites and 1760 terms. Prior to clustering, vectors representing sites were normalized to unit length. No further normalization or term weighting was applied. Following initial experiments on the data \cite{greene10dynak}, we set $k_{t}=20$ to identify high-level topical clusters, and used a balance parameter of $\alpha=0.1$ to allow a contribution from historic data, without overly constraining the co-clustering process. We examined a range of low matching thresholds $\theta \in [0.1,0.3]$. The general content of the resulting timelines was not significantly different for these values.  As representative results, we show the timelines identified for $\theta=0.2$.

\begin{figure}[!b]
\begin{center}
\includegraphics[width=4.2in]{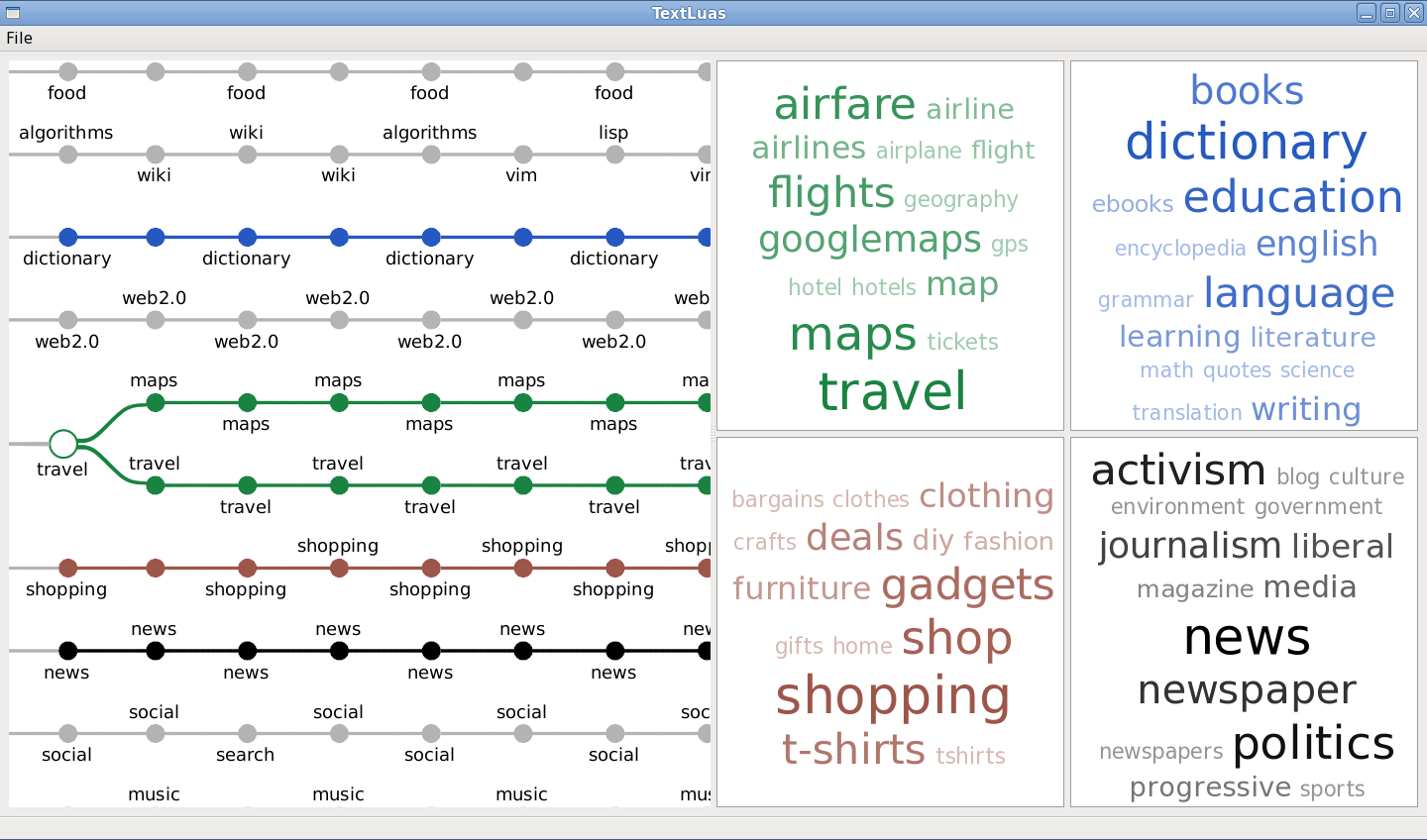}
\caption{\label{fig:del1}A subset of dynamic cluster timelines identified on the \emph{Del.icio.us} data set. The left-hand side shows the metro map view of the timelines over an 8 week period. The right-hand side shows aggregate tag clouds for the four selected paths.}
\end{center}
\end{figure}

\begin{figure}
\centering
\subfigure[``Maps'' timeline]{\includegraphics[width=3.4in]{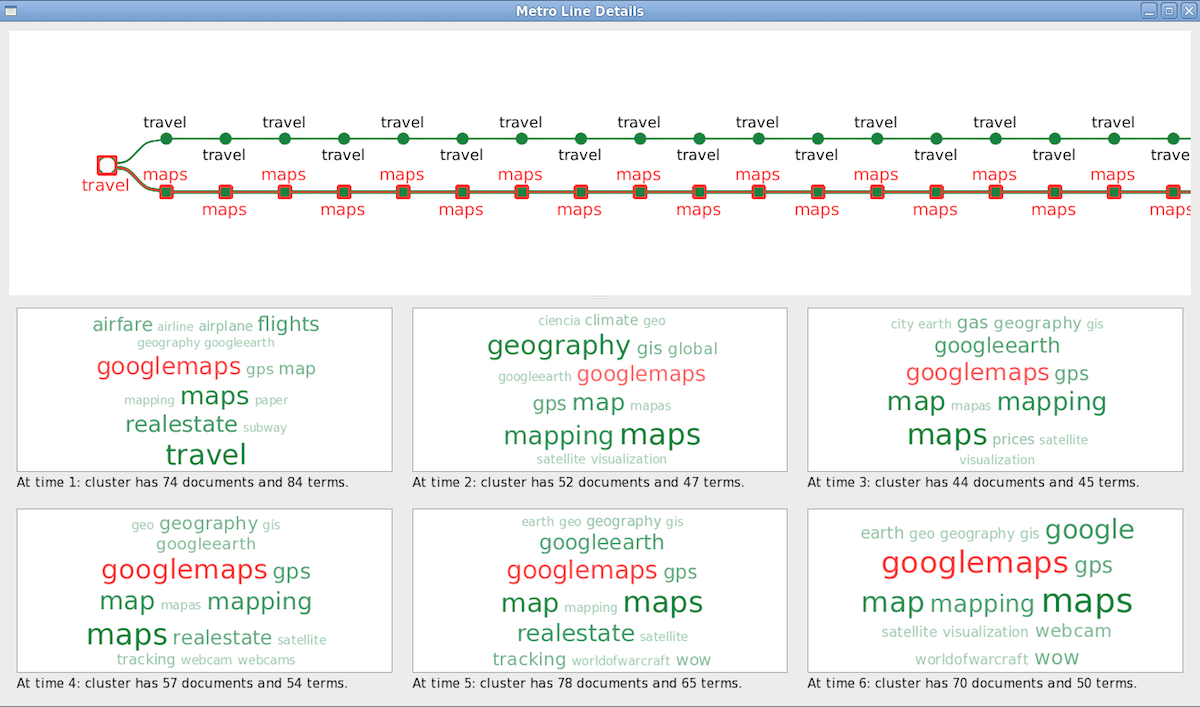}\label{fig:del2maps}}
\vskip 1em
\subfigure[``Travel'' timeline]{\includegraphics[width=3.4in]{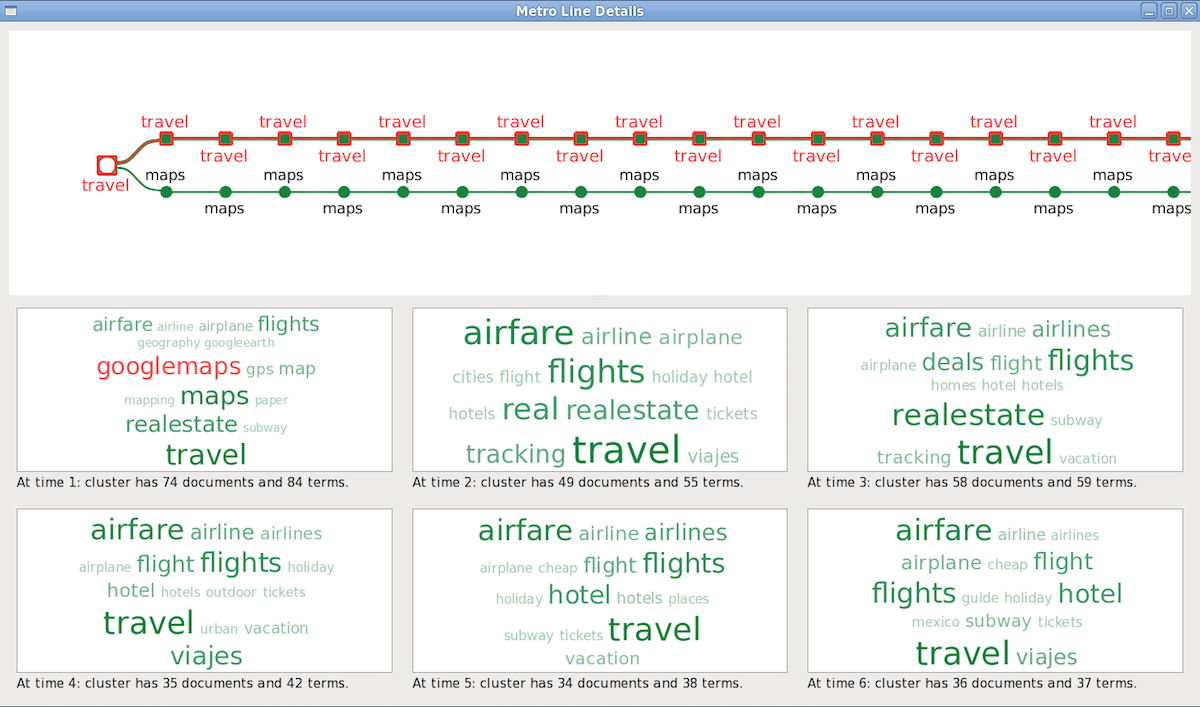}\label{fig:del2travel}}
\caption{\label{fig:del2}TextLuas metro line details view in the \emph{Del.icio.us} data set.  By pressing the arrow keys, we can alternate between the dynamic cluster above and below the split.  The top line contains topics on mapping \& geography while the bottom contains topics on travel.  Linked highlighting between the station clouds indicates that the term \textit {googlemaps} is a popular tag in the ``Maps'' timeline but is not present in the ``Travel'' timeline at all, except for at the split event.}
\end {figure}

When applied, the TextLuas metro map view shows us that the \emph{Del.icio.us} data set contains relatively distinct topics that persist through time. Dynamic clusters are generally born at $t=1$ (January 2006) and continue until $t=44$ (November 2006). This property is perhaps unsurprising, given that the data pertains to popular websites and tags -- we would expect that user-selected terms like ``shopping'' and ``travel'' will continue to be popular over time. \reffig{fig:del1} shows a number of the most persistent timelines. On the left-hand side, we see the timelines, zoomed to focus on a two-month period. The largely disjoint nature of the timelines is apparent. However, we did observe a small number of other life-cycle events in the data. For instance, there is a divergence in the dynamic cluster labeled ``travel.'' This split creates two timelines: one focusing on travel \& airlines and the other on mapping \& geography. We can use the metro line details view described in \refsec{secMetroDetail} to view details of this split and its two resulting dynamic clusters. Figures \ref{fig:del2maps} and \ref{fig:del2travel} show the difference in the content between the two timelines. The content of the station clouds suggests that the split appears to correspond to the increasing popularity of Google Maps during early 2006.  Linked highlighting between station tag clouds indicates that the keyword \textit {googlemaps} is consistently popular in the ``Maps'' timeline but only appears in the splitting station of the ``Travel'' timeline, illustrating the shift in topic.

\begin{figure}
\begin{center}
\includegraphics[width=3.4in]{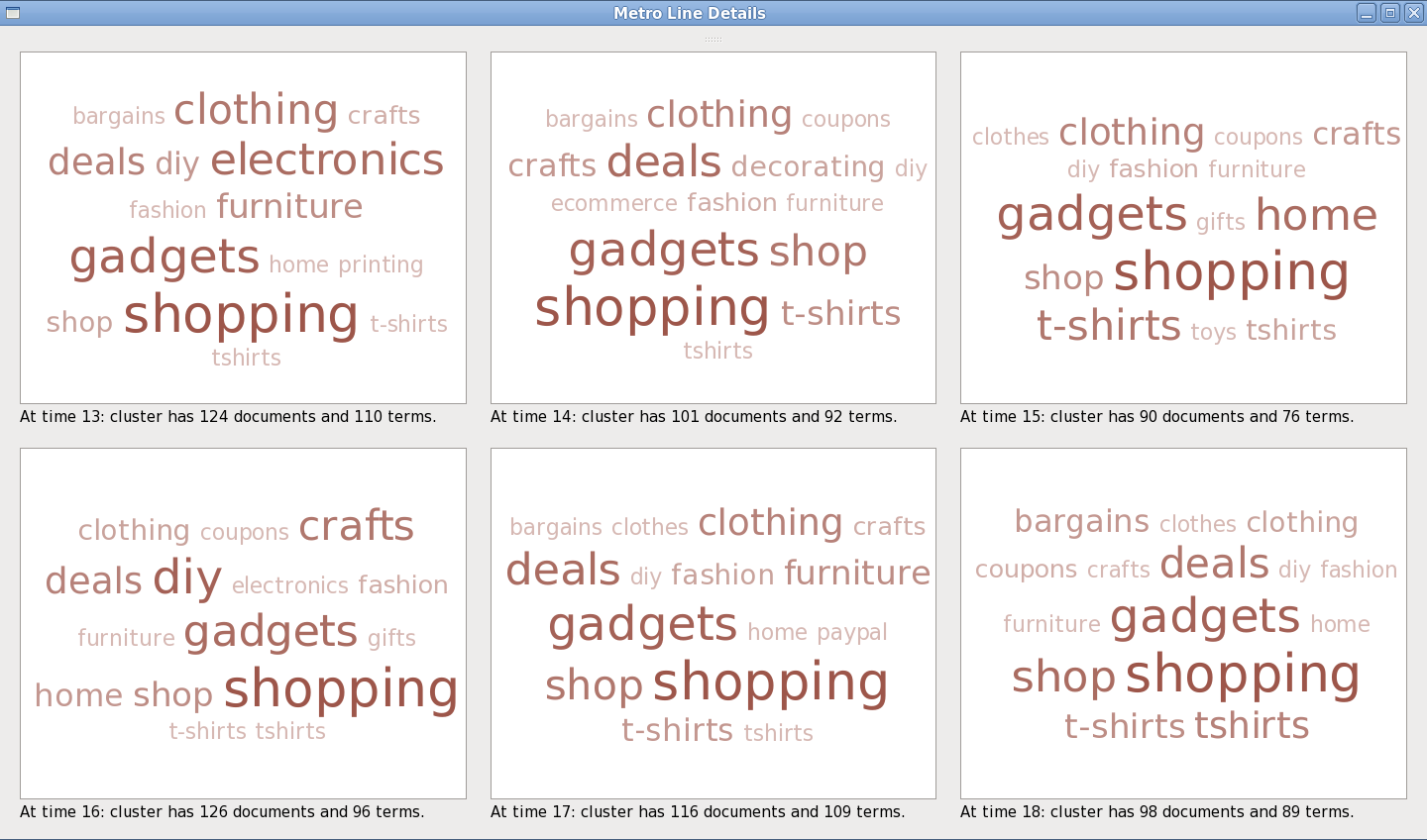}
\caption{\label{fig:del3}A station cloud matrix of metro line details, representing a dynamic cluster related to ``shopping'' in the \emph{Del.icio.us} data set. The tag clouds illustrate the changing content of the dynamic cluster over six time steps.}
\end{center}
\end{figure}

Although the station labels in \reffig{fig:del1} appear relatively static, and the majority of dynamic clusters continue uninterrupted from birth to death, we would expect there to be constant smaller changes in the popularity of bookmarks and terms between steps. To illustrate this fact, we again consult the metro line details view. In \reffig{fig:del3} we show the station cloud matrix for a dynamic cluster related to ``shopping''. Although the cluster's timeline has the same station label for all time steps in the metro map view, the clouds, shown in  \reffig{fig:del1}, provide a far more detailed representation of the content of each cluster.  For instance, we can see the changing prominence given to websites related to electronics, clothing, and furniture. 
 
\subsection{Economic News Data}
\label{sec:evalreal2}

\begin{figure}
\begin{center}
\includegraphics[width=4.5in]{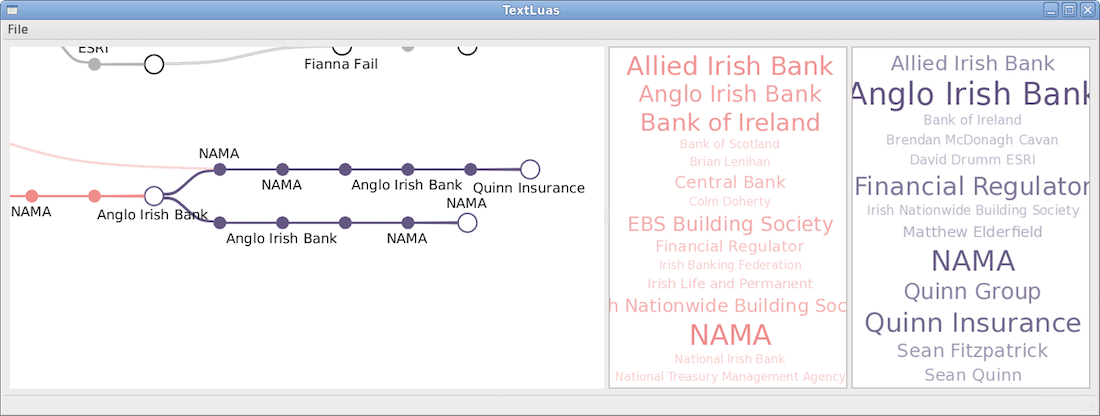}
\caption{\label{fig:irish1}Metro map view of a number of timelines identified on the Irish economic news data set. Several cluster life-cycle events are apparent:  continuation, merging, splitting, and intermittent behavior.}
\end{center}
\end{figure}
\begin{figure}
\begin{center}
\includegraphics[width=4.5in]{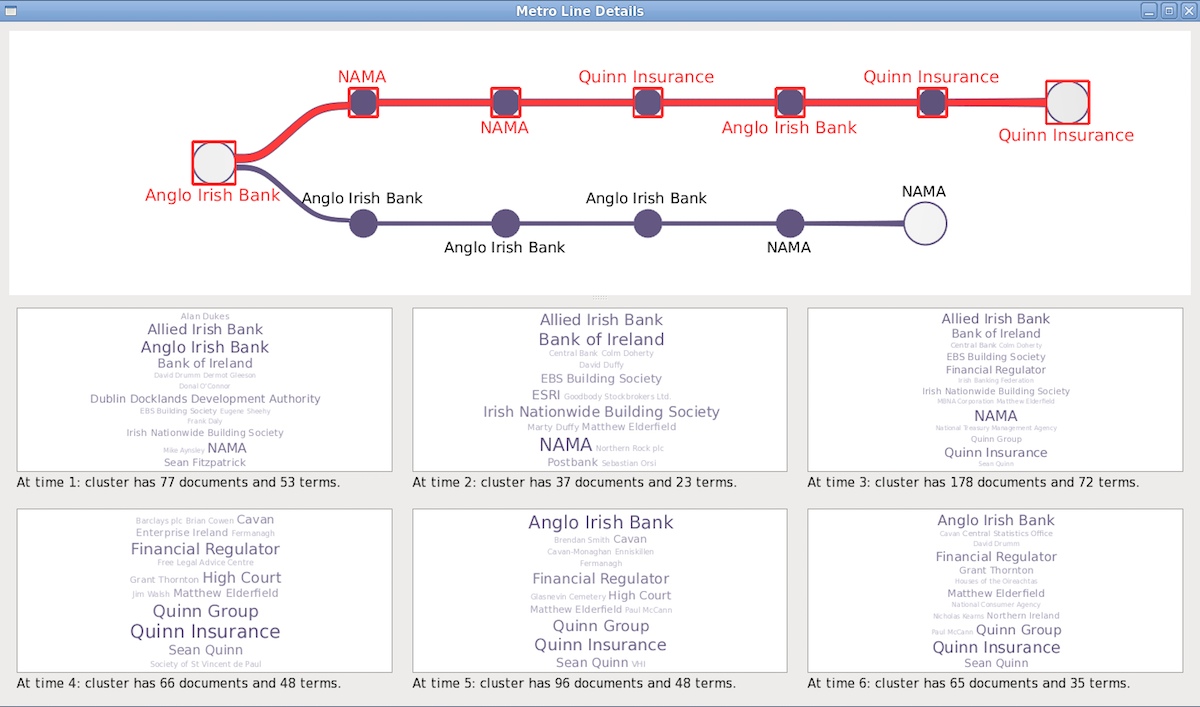}
\caption{\label{fig:irish2}Metro line details view for a split event related to the banking crisis as identified in the Irish economic news data set. The split corresponds to a development in the topic where Anglo Irish Bank proposed rescue plans for Quinn Insurance.}
\end{center}
\end{figure}

In our second case study, we analyze a corpus of articles relating to economic news from three online news sources (RTE, The Irish Times, The Irish Independent), which were previously collected for the purpose of sentiment classification \cite{brew10pais}. We use a set of 21,746 news articles covering a period from August 2009 to April 2010. The goal of this analysis is to uncover the dominant news stories and topics that pertained to the Irish and global economic situation during that time period. To represent each article, we extract a set of terms corresponding to the named entities in the corpus:  people, organizations, and geographical locations.  These entities were derived from a manually curated list of 16,265 entities extracted from The Irish Independent website\footnote{\url{http://www.independent.ie}}. The complete data set was divided into 36 time steps, each one week in duration. Each step graph contained approximately 604 articles described by 597 terms (entities).  To pre-process the data, terms appearing in less than two articles in a given step were removed, and article vectors were normalized to unit length. Given the large number of significant economic news stories reported on a weekly basis during the time the data was collected, we selected $k_{t}=25$ to identify step clusters. We used the same $\alpha$ parameter and range for $\theta$ as employed in \refsec{sec:evalreal1}. We observed that matching parameter values $\theta > 0.15$ occasionally lead to fragmentation of related topics.  For the results shown here, we use $\theta=0.15$.

In contrast to the previous data set, the TextLuas metro map reveals a more complex set of timeline structures and interactions for the news data. Again, this is expected, given the volatile nature of developing news stories. Many dynamic cluster life-cycle events are highlighted in the results. An interesting example is shown in \reffig{fig:irish1}, where the highlighted timelines reflect the coverage of the Irish banking crisis in the mainstream media. A blue timeline on the left, but mostly hidden from view, covers the Irish banking sector and the Irish ``bad bank'' NAMA intermittently appears for 30 time steps. We observe a split in this timeline at the first red node, reflecting a breaking news story in early April 2010 when an Irish bank proposed rescue plans for an insurance provider.  We can use the details view as shown in \reffig{fig:irish2} to examine the step clusters. The six station tag clouds provide a synopsis of the development of the story, highlighting key individuals and organizations that were involved. 

\begin{figure}
\begin{center}
\includegraphics[width=4.5in]{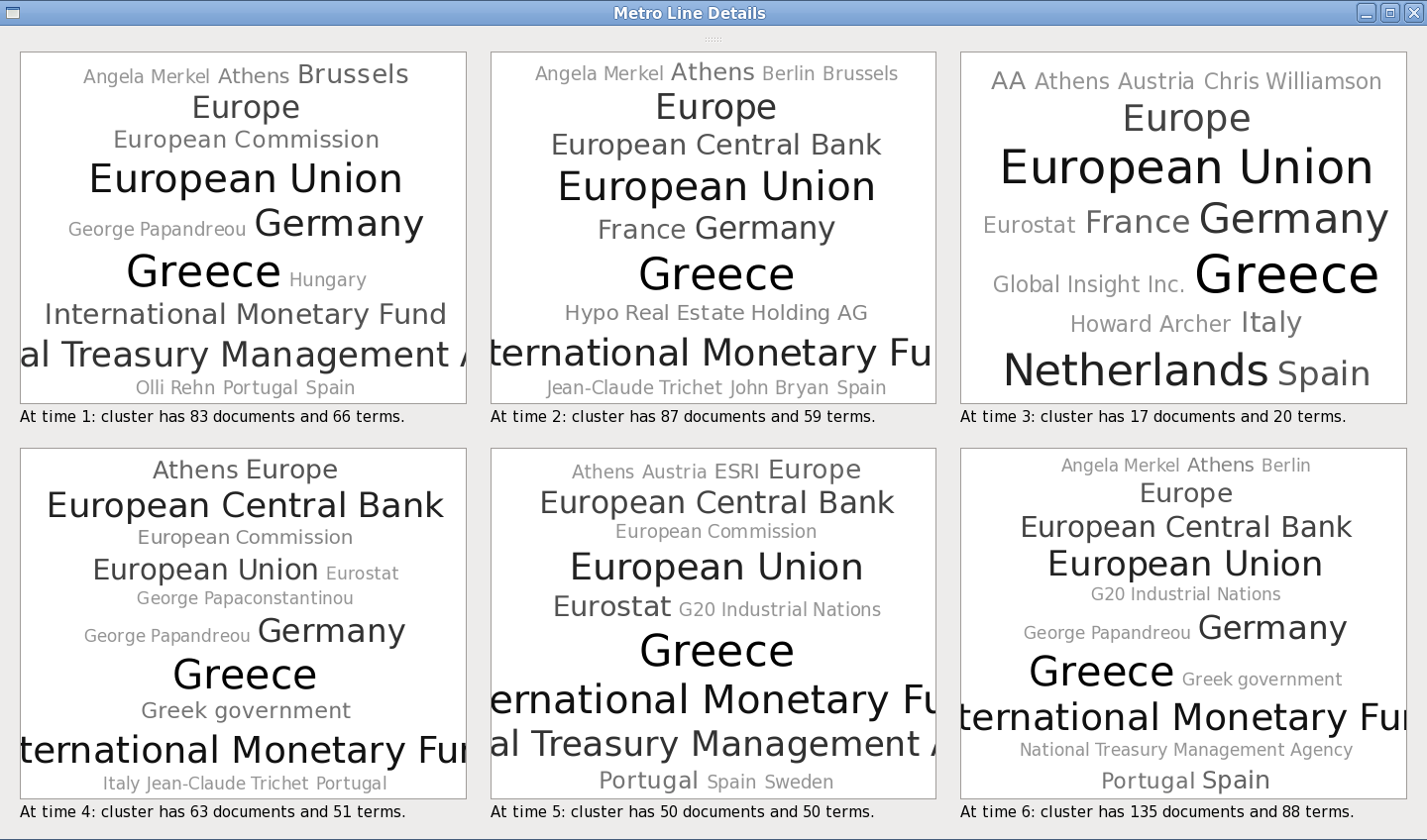}
\caption{\label{fig:irish3}A station cloud matrix from the TextLuas details view, highlighting the change in content of a dynamic cluster tracking the Greek economic situation over a six week period in 2010.}
\end{center}
\end{figure}

Another prominent news theme identified by the dynamic clustering process concerns the Greek economic situation in 2010. This dynamic cluster persisted consistently over 14 time steps, with the label ``Greece'' assigned to 13 of the stations. Naturally, there were a number of major developments related to the story during that period. To explore these in more detail, we can use the station cloud matrix. \reffig{fig:irish3} shows the clouds corresponding to the last six weeks of the timeline. From the changes in tag weightings, we can see the varying involvement of the EU, IMF, EU member states including Germany, and EU politicians, reflecting the news coverage of this crisis in March-April of 2010.

\subsection{User Feedback}
\label{sec:UserFeedback}
Recently, TextLuas has been used to illustrate evolution of scientific communities. In this case, the data set consists of a co-citation network of 5,772 researchers in two related disciplines of computer science -- Semantic Web~(SW) and Information Retrieval~(IR). In addition, 282,055 terms were extracted from their publications. This network was divided into ten step graphs spanning the period between 2000 and 2009 -- on average each step graph had 2,027 authors and 81,764 edges. This dynamic graph can be divided into scientific communities based on co-citation links. Specifically, the Louvain modularity optimization method \cite{Blondel2008} was used to determine the communities in this study. The community tracking was done by a ``best-match'' strategy, \ie the step clusters were connected by the highest Jaccard coefficient (\reft{eqn:jac}) of sets of authors, provided that the absolute overlap between the two step clusters was at least five authors. A more detailed description of the data set and the methodology can be found in~\cite{10Belak}.

Frequently, scientific communities exchange members and furthermore, themes associated with each community can shift or propagate to other communities. Research communities can split into several different communities or conversely, many can merge into a single community.  Many of the interesting cross-community effects are associated with simultaneous changes of the topics discussed in a community and its structure, \eg a sudden emergence of a community combining topics of different disciplines may be a start of a trans-disciplinary community.  When analyzing such phenomena in large dynamic networks, proper visualization and filtering techniques play a key role. The main objective for our collaborator was to observe community evolution, along with the typical life cycles in the data, and possible topic shifts in the literature that may occur. 

\begin{figure}
\begin{center}
\includegraphics[width=4.5in]{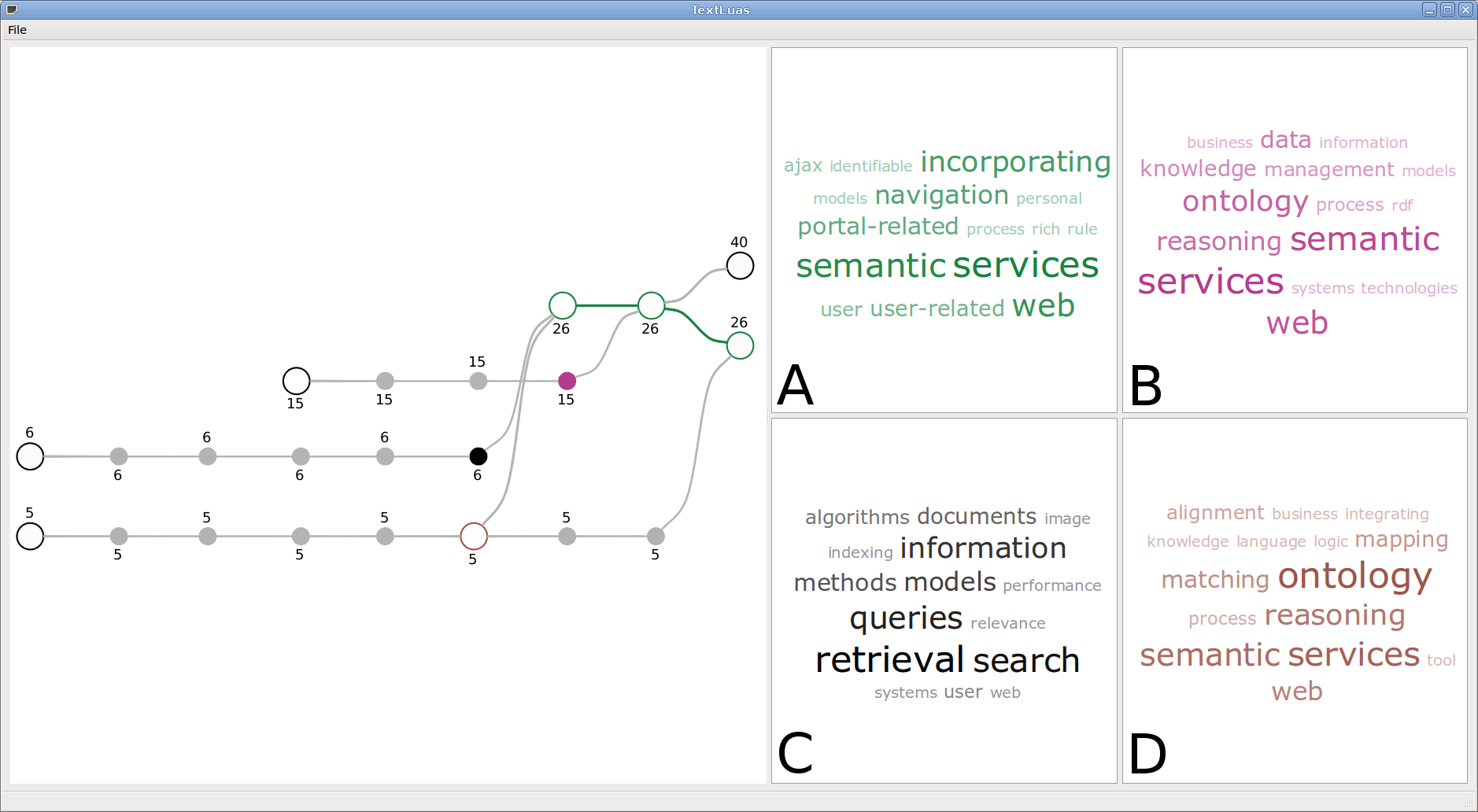}
\caption{\label{fig:communityInteraction}Diagram from Bel\'ak \textit{et al.} which illustrates the interaction between the main ancestors of community 26 (\textbf{A}): 15 ``semantic web'' (\textbf{B}), 6 ``web IR'' (\textbf{C}), 5 ``semantic web'' (\textbf{D}).}
\end{center}
\end{figure}

Prior to TextLuas, the researcher used the semi-automatic generation of life cycle diagrams produced using Graphviz\footnote{\url{http://www.graphviz.org}} and manual inspection of several large text files containing statistics about community overlaps, their characteristic themes, and other specifically-tailored measures for the analysis of cross-community phenomena. The latter tools were designed in order to understand the structure of the overlaps that existed between communities and identify the communities that caused the event. This methodology involved iterative formulation of hypotheses explaining the cross-community dynamics and successive generation of the appropriate life cycle diagrams. This process is difficult when many dynamic communities participated in a particular event. In order to publish results, the diagrams were redrawn by hand and tag clouds were generated separately in \LaTeX. Examples of these figures can be found in the older publication~\cite{10Belak}. 

\begin{figure}
\begin{center}
\includegraphics[width=4.5in]{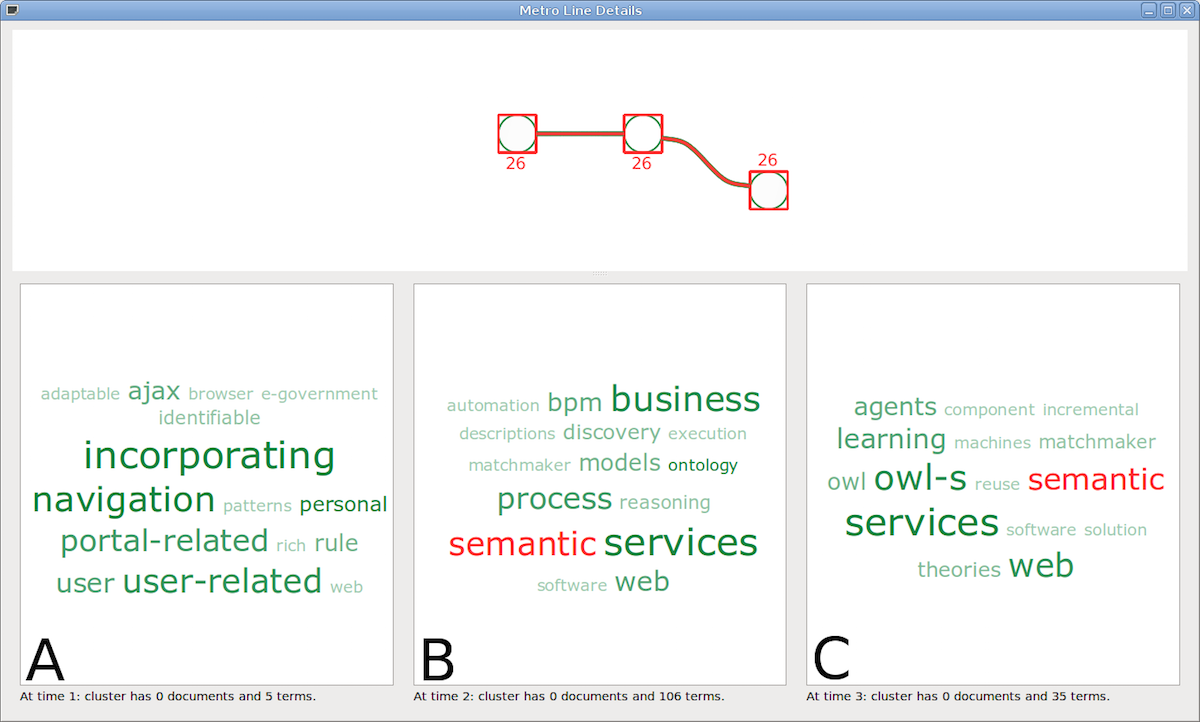}
\caption{\label{fig:communityInteractionDetail}Diagram from Bel\'ak \textit {et al.}, which illustrates a topic shift in community 26 from navigation-related topics towards semantic web (compare the tag cloud \textbf{A} with \textbf{B} and \textbf{C}). This corresponds with the influence of community 15.}
\end{center}
\end{figure}

As an alternative to the manual inspection and illustration of dynamic cluster structures, Figures~\ref{fig:communityInteraction}~and~\ref {fig:communityInteractionDetail} illustrate findings by this researcher visualized by TextLuas, as discussed \cite{11Belak}.  A custom names file was specified for the station
names based on the requirements for the user's tasks.
Figure~\ref{fig:communityInteraction} illustrates the emergence of community labeled 26 (tag cloud \textbf{A}), which has been identified as a potentially interesting as it was formed by distinct communities from both SW and IR fields. Moreover, the strong change of topics was subsequently identified in a second step of the community. Our collaborator wanted to examine the communities which had a significant impact on community 26, which could potentially shed some light on these dynamics. The stations that precede community 26 indicate that it was formed by SW community 5 and IR community 6 as seen by the terms in their corresponding tag clouds \textbf{D} and \textbf{C}, whereas topics like ``navigation'', ``personalization'' and ``web'' suggests interdisciplinary topics of community 26. The sudden change in the second station as seen in Figure ~\ref {fig:communityInteractionDetail} naturally leads to the question: What had happened? Was the community 26 influenced heavily by community 15, as suggested by a previous analytics, or was it just a false-positive and the community was rather influenced by completely new authors entering the field? In this case, the  researcher can say that the topic shift was most likely due to community 15, also depicted in Figure~\ref{fig:communityInteraction}, as tag cloud \textbf{B} flows into the second station of community 26. At this point, displaying community content through tag clouds was useful as it allowed the researcher to hypothesize about the causes of the cross-community events, \eg change of the topics, and immediately test the hypotheses. Compared to the previous manual methodology, the use of TextLuas, which places the content \textit{and} life cycle information in the same context, provided a faster and more efficient way to inspect the evolution of bibliographical communities. 

\section{Conclusion}
\label{sec:conc}
Currently, TextLuas makes heavy use of linked views to
show cluster content and step cluster evolution.  By linked views,
we mean that cluster evolution and tag cloud data is in two separate views,
but they are linked by using the same color and an update to one
influences the other.  In future work it may be possible to integrate both tag
information and cluster evolution into a single view, but further 
investigation is required.


TextLuas, in some senses, is scalable to large document collections.  In
our case studies, our visualization system scaled to collections of hundreds of
thousands of time-stamped documents.  However, currently TextLuas does
have a limitation in that it is not as scalable when the number of timeline 
events increases significantly (\eg highly-volatile dynamic graphs with many cluster splits, merges, births, and deaths).  Additionally, as it uses color
for linked highlighting, only a small number of timelines can be investigated
simultaneously.  In future work, we
would like to look at forms of dynamic cluster aggregation or filtering that would be 
appropriate for the task of visualizing group evolution in highly volatile data
and possible ways of increasing the number of timelines that can be compared
at once. To extend the scalability of the clustering at individual time steps, problem decomposition methods could potentially be applied \cite{anand09kais}.

In this paper, we presented TextLuas:  a system for identifying and visualizing dynamic clusters.  The contributions included a model for tracking persistent clusters across time and a system for visualizing the evolution of those dynamic clusters, in terms of both their structure \textit{and} their associated textual content. We applied the system on two data sets, one Web 2.0 social bookmarking data set and then a collection of economic news articles, 
and visualized the results from the perspective of life cycle events and cluster contents. We also solicited user
feedback from researchers interested in tracking the evolution of scientific communities via bibliographic network
analysis. This application suggested that TextLuas was a faster and more efficient way to inspect the dynamics of
scientific communities, which is cumbersome if the information pertaining to both the content and life cycles of clusters is not
integrated.


\section*{Acknowledgment}
\noindent This publication has emanated from research conducted with the financial support of Science Foundation Ireland (SFI) under Grant Numbers 08/SRC/I140 and SFI/12/RC/2289.

\bibliographystyle{agsm}
\bibliography{textluas,textluas-viz}

\end{document}